\documentclass[12pt,reqno]{amsart}

\usepackage{amsmath,amsfonts,amsthm,amssymb,amsxtra}
\usepackage{bbm} % to get \1
\usepackage[hidelinks]{hyperref} % creates hyperlinks to the references 

\newtheorem{theorem}{Theorem}%[section]

\newtheorem{lemma}[theorem]{Lemma}

\theoremstyle{definition}

\theoremstyle{remark}

\newtheorem{remark}[theorem]{Remark}

\newcommand{\1}{\mathbbm{1}}

\renewcommand{\epsilon}{\varepsilon}

\newcommand{\N}{\mathbb{N}}

\renewcommand{\phi}{\varphi}
\newcommand{\R}{\mathbb{R}}

\newcommand{\Sph}{\mathbb{S}}

\DeclareMathOperator{\Tr}{Tr}

\begin{document}

\title[A Jensen inequality for partial traces]{A Jensen inequality for partial traces and applications to partially semiclassical limits}

\author{Eric A.~Carlen}
\address[Eric A.~Carlen]{Department of Mathematics, Hill Center,
	Rutgers University, 110 Frelinghuysen Road, Piscataway NJ 08854-8019, USA}
\email{carlen@math.rutgers.edu}

\author{Rupert L.~Frank}
\address[Rupert L.~Frank]{Mathe\-matisches Institut, Ludwig-Maximilians Universit\"at M\"unchen, The\-resienstr.~39, 80333 M\"unchen, Germany, and Munich Center for Quantum Science and Technology, Schel\-ling\-str.~4, 80799 M\"unchen, Germany, and Mathematics 253-37, Caltech, Pasa\-de\-na, CA 91125, USA}
\email{r.frank@lmu.de}

\author{Simon Larson}
\address[Simon Larson]{Mathematical Sciences, Chalmers University of Technology and the University of Gothenburg, SE-41296 Gothenburg, Sweden}
\email{larsons@chalmers.se}

\thanks{\copyright\, 2025 by the authors. This paper may be reproduced, in its entirety, for non-commercial purposes.\\
	Partial support through US National Science Foundation grant DMS-1954995 (R.L.F.), the German Research Foundation grants EXC-2111-390814868 and TRR 352-Project-ID 470903074 (R.L.F.), as well as the Swedish Research Council grant no.~2023-03985 (S.L.) is acknowledged.}

\begin{abstract}
	We prove a matrix inequality for convex functions of a Hermitian matrix on a bipartite space. As an application we reprove and extend some theorems about eigenvalue asymptotics of Schr\"odinger operators with homogeneous potentials. The case of main interest is where the Weyl expression is infinite and a partially semiclassical limit occurs.
\end{abstract}

\maketitle

\section{Introduction and main results}

\subsection{A Jensen inequality partial traces}

A simple, yet very useful inequality says that if $H$ is a Hermitian matrix in a finite-dimensional Hilbert space $\mathcal H$ and $f$ is a convex function defined on the convex hull of the spectrum of $H$, then for any normalized $\psi\in\mathcal H$
\begin{equation}
	\label{eq:jensenintro}
	f\left( \langle \psi| H | \psi \rangle \right) \leq \langle \psi | f(H) | \psi \rangle \,.
\end{equation}
This well-known result easily follows from Jensen's inequality, applied to the spectral measure of $H$; see, e.g., \cite[Proof of Theorem 2.9]{Car1} or \cite[Lemma 3.2]{Car2}. 

Our goal in this paper is to extend this inequality to the bipartite setting where $\mathcal H=\mathcal H_1\otimes\mathcal H_2$ is the tensor product of two spaces. As usual we denote by $\Tr_j$, $j=1,2$, the partial traces. For background on these matters we refer to \cite[Section 5]{Car1} and \cite[Chapter 2]{Car2}. The extension is motivated by a specific application that we also discuss here. 

The inequality that we will prove says that for any normalized $\phi\in\mathcal H_1$, and selfadjoint $H$ on $\mathcal H_1\otimes\mathcal H_2$
\begin{equation}
	\label{eq:jensenbipartintro}
	\Tr_2 f(\langle \phi|H|\phi\rangle) \leq \langle \phi|{\Tr_2 f(H)}| \phi \rangle \, ,
\end{equation}
where $\langle \phi|H|\phi\rangle$ on the left side denotes the  operator $\Tr_1[(|\phi\rangle\langle \phi|\otimes \1_{\mathcal H_2}) H]$  on $\mathcal H_2$. Clearly, when the space $\mathcal H_2$ is trivial, inequality \eqref{eq:jensenbipartintro} reduces to \eqref{eq:jensenintro}.

In fact, we will prove the following extension of \eqref{eq:jensenbipartintro}, where $|\phi\rangle\langle\phi|$ is replaced by a density matrix (that is, a nonnegative operator of unit trace).

\begin{theorem}
	\label{mainjensen}
	Let $\mathcal H_1$, $\mathcal H_2$ be finite dimensional Hilbert spaces, let $H$ be a Hermitian matrix in $\mathcal H_1\otimes\mathcal H_2$ and let $f$ be a convex function on the convex hull of the spectrum of $H$. Then for any density matrix $\rho$ on $\mathcal H_1$
	$$
	\Tr_2 f( \Tr_1 (\rho\otimes\1)^\frac12 H (\rho\otimes\1)^\frac12) \leq \Tr_1 \rho^\frac12 \left( \Tr_2 f(H) \right) \rho^\frac12 \,.
	$$
\end{theorem}

We will prove this theorem in Section \ref{sec:mainjensen}.

%%%%%%%%%%%%%%%%%%%%%%%%%%%%%%

\subsection{Partially semiclassical limits}

Our interest in inequality \eqref{eq:jensenbipartintro} comes from what we call a \emph{partially semiclassical limit} and from three recent papers, discussed below, where this limit appears naturally in applications. We are concerned with the asymptotic behavior of eigenvalues of differential operators. The leading term in these asymptotics is often given by Weyl's law, but in our applications this expression for the leading term given by Weyl's law is infinite. In some situations where this happens, an asymptotic separation of variables occurs. For one group of variables Weyl's law is applicable and these variables become `semiclassical', while the complementary set of variables remains `quantum', that is, there appear differential operators that act with respect to the `quantum variables' and depend parametrically on the `semiclassical variables'. We call this phenomenon a `partially semiclassical limit' and give more references where this is studied later on in this introduction. 

The description may seem vague at this point, but we hope it becomes clearer after stating Theorems \ref{warmup} and \ref{main}. We emphasize that these theorems are known, at least under certain additional regularity assumptions, and that our goal is to provide simple proofs for them, in the spirit of works of Berezin \cite{Be} and Lieb \cite{Li}, based on inequalities \eqref{eq:jensenintro} and \eqref{eq:jensenbipartintro}.

Both theorems concern Schr\"odinger operators
$$
H = -\Delta +V
\qquad\text{in}\ L^2(\R^d)
$$
with potentials $V\geq 0$ that are homogeneous of positive degree. More specifically, we are interested in the asymptotic growth as $\lambda\to\infty$ of the number $N(\lambda,H)$ of eigenvalues $<\lambda$, counting multiplicities. The following constant appears in the limits,
$$
C_{\gamma,d} := (4\pi)^{-\frac d2}\, \gamma^{-1} \frac{\Gamma(\frac d\gamma)}{\Gamma(\frac d\gamma+\frac d2 +1)} \,.
$$

The first theorem, which we state as a warm-up, involves a standard semiclassical limit.

\begin{theorem}\label{warmup}
	Let $d\in\N$ and $\gamma> 0$. Let $0\leq V\in L^1_{\rm loc}(\R^d)$ be homogeneous of degree~$\gamma$. Then
	$$
	\lim_{\lambda\to\infty} \lambda^{-\frac{d(\gamma+2)}{2\gamma}}
	N(\lambda,H) = 
	C_{\gamma,d} \int_{\Sph^{d-1}} V(\omega)^{-\frac d\gamma}\,d\omega \,.
	$$	
\end{theorem}

We emphasize that this theorem is valid whether or not the integral on the right side is finite. In the next theorem, we consider a situation where it is infinite (see Remark~\ref{infiniterem} below), which gives rise to a partially semiclassical limit. We write $d=m+n$, $\gamma=\alpha+\beta$ and denote coordinates in $\R^{m+n}$ by $(x,y)\in\R^m\times\R^n$.

\begin{theorem}\label{main}
	Let $n,m\in\N$ and $\alpha,\beta>0$ with $m\alpha^{-1}>n\beta^{-1}$. Let $0\leq V\in L^1_{\rm loc}(\R^{n+m})$ be separately homogeneous of degrees $\alpha$ and $\beta$ with respect to $x$ and $y$, respectively. Then
	$$
	\lim_{\lambda\to\infty} \lambda^{-\frac{m(\alpha+\beta+2)}{2\alpha}} N(\lambda,H) = 
	C_{\frac{2\alpha}{\beta+2},m} \int_{\Sph^{m-1}}  \Tr \Bigl( (-\Delta_{y'} + V(\omega,y'))^{-\frac{m(\beta+2)}{2\alpha}} \Bigr)\,d\omega\, .
	$$
\end{theorem}

As before, the theorem is valid whether or not the integral on the right side is finite. Also, a similar theorem holds when $m\alpha^{-1}<n\beta^{-1}$ by switching the roles of $x$ and $y$. One can also compute the asymptotics when $m\alpha^{-1}=n\beta^{-1}$, but they do not involve a partially semiclassical limit; see the references given below.

Theorem \ref{warmup} describes a semiclassical limit, since the leading term
\begin{equation}
	\label{eq:sccalc1}
	C_{\gamma,d}\ \lambda^{\frac{d(\gamma+2)}{2\gamma}} \int_{\Sph^{d-1}}  V(\omega)^{-\frac d\gamma}\,d\omega = \iint_{\R^d\times\R^d}   \1( |\xi|^2 + V(x)<\lambda)\,\frac{dx\,d\xi}{(2\pi)^d}
\end{equation}
is given by an integral over semiclassical phase space. In contrast Theorem \ref{main} describes a partially semiclassical limit, since the leading term
\begin{equation}\label{eq:sccalc2}
\begin{aligned}
	& C_{\frac{2\alpha}{\beta+2},m} \lambda^{\frac{m(\alpha+\beta+2)}{2\alpha}} \int_{\Sph^{m-1}}  \Tr \Bigl( (-\Delta_{y'} + V(\omega,y'))^{-\frac{m(\beta+2)}{2\alpha}} \Bigr)\,d\omega \\
	& = \iint_{\R^m\times\R^m}    N(\lambda,|\xi|^2 -\Delta_y + V(x,y))\,\frac{dx\,d\xi}{(2\pi)^m}
\end{aligned}
\end{equation}
is given by an integral over part of the semiclassical phase space, namely $\R^m\times\R^m$. Associated to each given $(x,\xi)\in\R^m\times\R^m$ is an effective Schr\"odinger operator $|\xi|^2 -\Delta_y +V(x,y)$ in $L^2(\R^n)$, and the limit depends on the spectrum of these operators. 

The proof of identities \eqref{eq:sccalc1} and \eqref{eq:sccalc2} follows by straightforward computations with beta functions, using also the explicit expression for the volume of the unit ball. Similar computations appear in the proofs of Theorems \ref{warmupheat} and \ref{mainheat} below and here we omit the details.

The  partially semiclassical limit  phenomenon has been studied since the early 1960's and we refer to \cite[Chapter~5, Section: Commentary and references to the literature]{BiSo} for many references. Those include, in particular, results by Solomyak and Vulis concerning a power-like degeneration of the coefficients of an operator close to the boundary of a domain; see \cite[Theorem 5.19]{BiSo}. See also \cite{Ta}. The same phenomenon in the setting of Schr\"odinger operators was studied by Robert \cite{Ro} and by Simon \cite{Si}. The latter studied the operators $-\Delta +|x|^\alpha |y|^\beta$ in $L^2(\R^2)$ with $\alpha,\beta>0$, which is a special case of Theorem \ref{main} with $m=n=1$. In Solomyak's paper \cite{So}, Theorems \ref{warmup} and \ref{main} appear under certain additional technical assumptions (continuity of $V$ in both cases; $d\geq 3$ and finiteness of the integral in Theorem \ref{main}; nonvanishing of $V$ outside the set $(\R^m\times\{0\})\cup(\{0\}\times\R^n)$). Simon \cite{Si} and Solomyak \cite{So} also study the case $m\alpha^{-1} = n\beta^{-1}$. 

The three recent papers that motivated us are \cite{Re,CdVdHDiTr,FrLa}. In \cite{Re} the author computes the asymptotic number of low-lying states in a two-dimensional confined Stark effect and finds an asymptotic separation of variables. In \cite{CdVdHDiTr} the authors compute the asymptotic growth of eigenvalues for manifolds whose metric degenerates near the boundary. In passing, we also mention the related paper \cite{DiRe} where techniques from \cite{Re} and \cite{CdVdHDiTr} are combined. In \cite{FrLa}, two of us computed the asymptotic number of eigenvalues of Laplace operators less than $\lambda_j$ on a sequence of convex bounded open sets $\Omega_j$ satisfying $\lambda_j^\frac d2|\Omega_j|\gg 1 \sim \lambda_j^\frac12 r_{\rm in}(\Omega_j)$. (Here $r_{\rm in}(\Omega_j)$ is the inradius of $\Omega_j$.) The results obtained in the present paper allow us to reprove the asymptotics in \cite{FrLa} in the case of Dirichlet boundary conditions, but do not appear to yield the results obtained in \cite{FrLa} for Neumann boundary conditions; the approach in  \cite{FrLa}, based on Dirichlet--Neumann bracketing, gives a unified proof.

We emphasize again that in the present paper, while we  remove some unnecessary assumptions from \cite{So}, we do not strive at obtaining the most general results. Rather, we aim to present an approach to partially semiclassical limits that maintains as close as possible a parallel with methods that yield semiclassical limits, and 
would like to present our method in the simplest possible setting. For this reason, we first present the method in the setting of Theorem \ref{warmup}, which only uses arguments that are already present in the semiclassical limit literature. Then we show how a natural extension of these arguments leads, through Theorem~\ref{mainjensen}, to Theorem \ref{main}.

Our proof is based on heat kernel asymptotics and coherent states. The idea of deducing asymptotics for $N(\lambda,H)$ as $\lambda\to\infty$ from asymptotics for $\Tr e^{-tH}$ as $t\to 0$ goes back to Carleman \cite{Ca} and is based on a Tauberian theorem. (More precisely, Carleman used the closely related resolvent trace asymptotics instead of heat trace asymptotics.) Denoting
$$
C_{\gamma,d}' := \Gamma(\tfrac d\gamma+\tfrac d2 +1) \, C_{\gamma,d} = (4\pi)^{-\frac d2}\, \gamma^{-1} \Gamma(\tfrac d\gamma)
$$
and recalling the Hardy--Littlewood--Karamata Tauberian theorem in the form \cite[Theorem 10.3]{Si0}, we see that Theorems \ref{warmup} and \ref{main} follow from (in fact, are equivalent to) the following two theorems.

\begin{theorem}\label{warmupheat}
	Let $H$ be as in Theorem \ref{warmup}. Then
	$$
	\lim_{t\to 0} t^{\frac{d(\gamma+2)}{2\gamma}} \Tr e^{-tH} = C_{\gamma,d}' \int_{\Sph^{d-1}}  V(\omega)^{-\frac d\gamma} \,d\omega\,.
	$$
\end{theorem}

\begin{theorem}\label{mainheat}
	Let $H$ be as in Theorem \ref{main}. Then
	$$
	\lim_{t\to 0} t^{\frac{m(\alpha+\beta+2)}{2\alpha}} \Tr e^{-tH} 
	= C_{\frac{2\alpha}{\beta+2},m} \int_{\Sph^{m-1}}  \Tr \Bigl( (-\Delta_{y'} + V(\omega,y'))^{-\frac{m(\beta+2)}{2\alpha}} \Bigr)\,d\omega\,.
	$$
\end{theorem}

To emphasize the (partially) semiclassical character of these asymptotics we note that we can write
$$
C_{\gamma,d}' \, t^{-\frac{d(\gamma+2)}{2\gamma}} \int_{\Sph^{d-1}}  V(\omega)^{\frac d\gamma} \,d\omega= \iint_{\R^d\times\R^d}   e^{-t(|\xi|^2 + V(x))}\frac{dx\,d\xi}{(2\pi)^d}
$$
and
\begin{align*}
	& C_{\frac{2\alpha}{\beta+2},m} \, t^{- \frac{m(\alpha+\beta+2)}{2\alpha}} \int_{\Sph^{m-1}}  \Tr \Bigl( (-\Delta_{y'} + V(\omega,y'))^{-\frac{m(\beta+2)}{2\alpha}} \Bigr)\,d\omega \\
	& = \iint_{\R^m\times\R^m}   \Tr_{L^2(\R^n)} \bigl(e^{-t(|\xi|^2-\Delta_y + V(x,y))}\bigr)\,\frac{dx\,d\xi}{(2\pi)^m} \,.
\end{align*}
These identities will be derived in the course of the proof of Theorems \ref{warmupheat} and \ref{mainheat}.

From now on we will focus on the proofs of the latter two theorems. The advantage of working with heat traces is that the `difficult' upper bound comes for free by means of the Golden--Thompson inequality. In the setting of Theorem \ref{warmup} this is the standard Golden--Thompson inequality, while in that of Theorem \ref{mainheat} it is a partial variant of it, noted by Simon in \cite{Si}.

As an aside we mention that we could also consider the asymptotics of $\Tr(H-\lambda)_-^\gamma$ for some $\gamma\geq\frac32$. On the one hand, by a Tauberian-type argument this would give the asymptotics of $N(H,\lambda)$. On the other hand, we could use the sharp Lieb--Thirring inequality (for operator-valued potentials \cite{LaWe} in the setting of Theorem \ref{main}) to obtain the `difficult' upper bound by the limiting expression. For a recent implementation of this idea in a special case see \cite{AlLa}.

Thus, the only thing that needs to be proved is the lower bound in Theorems \ref{warmupheat} and~\ref{mainheat}. As we will show, this can be accomplished rather easily using coherent states. It is for this purpose that we need \eqref{eq:jensenintro} in the proof of Theorem \ref{warmupheat}. Our new inequality \eqref{eq:jensenbipartintro} plays the analogous role in the proof of Theorem \ref{mainheat}.

This proof of the lower bound in Theorem \ref{mainheat} using coherent states differs from that of Simon who uses the Feynman--Kac formula. We hope that our proof retains some of the elegance of Simon's proof of the upper bound. The use of coherent states in the context of eigenvalue asymptotics goes back at least to the celebrated papers by Berezin \cite{Be} and Lieb \cite{Li}. The usefulness of this method is further explained in \cite{Si1,Li2}; see also \cite{Fr} for a recent application to Weyl laws for Schr\"odinger operators on domains under minimal assumptions on the potential.

%%%%%%%%%%%%%%%%%%%%%%%

\section{Proof of Theorem \ref{mainjensen}}\label{sec:mainjensen}

We work under the assumptions of Theorem \ref{mainjensen}, that is, let $\mathcal H_1,\mathcal H_2$ be finite dimensional Hilbert spaces, let $H$ be a Hermitian matrix in $\mathcal H_1\otimes\mathcal H_2$ and let $\rho$ be a density matrix on $\mathcal H_1$. We set
$$
K:= \Tr_1 (\rho\otimes\1)^\frac12 H (\rho\otimes\1)^\frac12
$$
and chose an orthonormal basis $(v_1,\ldots,v_N)$ of $\mathcal H_2$ consisting of eigenvectors of $K$. Then, for any convex function $f$ on the convex hull of the spectrum of $H$,
\begin{equation}
	\label{eq:mainjensenproof}
	\Tr_2 f(K) = \sum_{n=1}^N \langle v_n | f(K) | v_n \rangle = \sum_{n=1}^N f(\langle v_n|K| v_n\rangle) \,.
\end{equation}
Now we write
$$
\rho = \sum_{m=1}^M \lambda_m |u_m\rangle\langle u_m|
$$
with an orthonormal basis $(u_1,\ldots,u_M)$ of $\mathcal H_1$. Then, for any $v\in\mathcal H_2$, we have
\begin{align*}
	\langle v| K | v\rangle & = \sum_{m=1}^M \bigl\langle u_m \otimes v \bigl| (\rho\otimes\1)^\frac12 H (\rho\otimes\1)^\frac12 \bigr| u_m\otimes v \bigr\rangle \\
	& = \sum_{m=1}^M \lambda_m \left\langle u_m \otimes v \left| H \right| u_m\otimes v \right\rangle .
\end{align*}
We fix $n\in\{1,\ldots,N\}$ and apply this identity with $v=v_n$. Using Jensen's inequality twice, we find
\begin{align*}
	f(\langle v_n|K| v_n\rangle) & = f\left(  \sum_{m=1}^M \lambda_m \left\langle u_m \otimes v_n \left| H \right| u_m\otimes v_n \right\rangle \right) \\
	& \leq \sum_{m=1}^M \lambda_m \, f\left(\left\langle u_m \otimes v_n \left| H \right| u_m\otimes v_n \right\rangle\right) \\
	& \leq \sum_{m=1}^M \lambda_m \left\langle u_m \otimes v_n \left| f(H) \right| u_m\otimes v_n \right\rangle .
\end{align*}
Here, the first application of Jensen's inequality uses $\lambda_m\geq 0$ and $\sum_{m=1}^M \lambda_m = \Tr_1\rho=1$, while the second application is inequality \eqref{eq:jensenintro}. Summing this inequality with respect to $n$, interchanging the two sums and recalling \eqref{eq:mainjensenproof}, we obtain
\begin{align*}
	\Tr_2f(K) & \leq \sum_{m=1}^M \lambda_m \sum_{n=1}^N \left\langle u_m \otimes v_n \left| f(H) \right| u_m\otimes v_n \right\rangle \\
	& = \sum_{m=1}^M \lambda_m \langle u_m |{\Tr_2 f(H)}| u_m \rangle \\
	& = \Tr_1 \rho^\frac12 \left( \Tr_2 f(H) \right) \rho^\frac12 \,.
\end{align*}
This completes the proof of the claimed inequality.

\begin{remark}\label{mainjensenrem}
	We will need an extension of Theorem \ref{mainjensen} to the infinite-dimensional setting. We assume that the operator $H$ and the function $f$ are nonnegative and that the operator $\Tr_1 (\rho\otimes\1)^\frac12 H (\rho\otimes\1)^\frac12$ has discrete spectrum. Then the above proof goes through unchanged, except that now $N$ and/or $M$ are possibly infinite. Since all quantities are nonnegative under our assumptions, all manipulations are allowed even if some of the sums are infinite. There are more subtle extensions to the infinite dimensional context, but this one is good enough for our purposes.
\end{remark}

%%%%%%%%%%%%%%%%%%%%%%%

\section{Semiclassical limit: Proof of Theorem \ref{warmupheat}}

In this section we prove Theorem \ref{warmupheat} and thereby also Theorem \ref{warmup}. It serves as a warm-up for the next section and is included mostly for pedagogical purposes. In particular, we want to highlight the role of inequality \eqref{eq:jensenintro} in this proof, which in the next section will be replaced by the new inequality \eqref{eq:jensenbipartintro}.

We proceed by proving an upper and a lower bound on $\Tr e^{-tH}$. To do so, we argue similarly as in \cite{Si1}, but a different choice of coherent states will allow us to relax the assumptions on $V$ imposed there.

\medskip

The \emph{upper bound} follows immediately from the Golden--Thompson inequality; see, e.g., \cite[Theorem 4.49]{Car2}. Specifically,
\begin{align*}
	\Tr e^{-tH} \leq \Tr e^{t\Delta/2} e^{-tV} e^{t\Delta/2} = (4\pi t)^{-\frac d2} \int_{\R^d}  e^{-tV(x)} \,dx\,.
\end{align*}
Introducing spherical coordinates $x=r\omega$ with $r>0$, $\omega\in\Sph^{d-1}$ and then changing variables by letting $s=t r^\gamma$ we obtain
\begin{align*}
	\int_{\R^d}  e^{-tV(x)}\,dx & = \int_{\Sph^{d-1}}  \int_0^\infty   \, e^{-t r^\gamma V(\omega)} r^{d-1}\,dr \,d\omega \\
	& = \gamma^{-1} \, t^{-\frac d\gamma} \int_{\Sph^{d-1}} \int_0^\infty  e^{-sV(\omega)}  s^{\frac{d}{\gamma}-1}\,ds\,d\omega\\
	& =	
	\gamma^{-1} \, t^{-\frac d\gamma} \, \Gamma(\tfrac d\gamma) \int_{\Sph^{d-1}} V(\omega)^{-\frac d\gamma}\,d\omega \,.
\end{align*}
This proves the upper bound. To express it as  semiclassical bound, and connect it with the lower bound that follows, we note that
$$
(4\pi t)^{-\frac d2} \int_{\R^d} e^{-tV(x)}\,dx = \iint_{\R^d\times\R^d}  e^{-t(|\xi|^2 +V(x))}\,\frac{dx\,d\xi}{(2\pi)^d}  \,.
$$

\medskip

For the \emph{lower bound} we fix a symmetric decreasing function $g\in H^1(\R^d)\cap L^\infty(\R^d)$ with $\|g\|_{L^2(\R^d)} = 1$ and compact support, and we set, for $(x,\xi)\in\R^d\times\R^d$,
$$
\psi_{\xi,x}(x') := e^{i\xi\cdot x}\, g(x'-x) \,.
$$
Then, by a  well-known consequence of  Plancherel's theorem,
$$
\int_{\R^d\times\R^d}  |\psi_{\xi,x}\rangle \langle \psi_{\xi,x}|\,\frac{dx\,d\xi}{(2\pi)^d}  = \1_{L^2(\R^d)} \,.
$$
Thus,
\begin{align*}
	\Tr e^{-tH} = \iint_{\R^d\times\R^d}    \Tr \bigl(|\psi_{\xi,x}\rangle \langle \psi_{\xi,x}| e^{-tH}\bigr) \,\frac{dx\,d\xi}{(2\pi)^d}
\end{align*}
and, by Jensen's inequality \eqref{eq:jensenintro} (generalized to the infinite dimensional setting),
$$
\Tr \bigl(|\psi_{\xi,x}\rangle \langle \psi_{\xi,x}| e^{-tH}\bigr) \geq e^{-t \langle \psi_{\xi,x} | H | \psi_{\xi,x} \rangle} \,.
$$
Standard computations with coherent states (see, e.g, \cite[Chapter 12]{LiLo} or \cite{Si1,Fr}) imply that
$$
\langle \psi_{\xi,x} | H | \psi_{\xi,x} \rangle = |\xi|^2 + \|\nabla g\|_{L^2(\R^d)}^2 + g^2*V(x) \,.
$$
Thus, we have shown that
\begin{align*}
	\Tr e^{-tH} & \geq e^{-t\|\nabla g\|_{L^2(\R^d)}^2} \iint_{\R^d\times\R^d}   e^{-t(|\xi|^2 + g^2*V(x))} \,\frac{dx\,d\xi}{(2\pi)^d}\\
	& = (4\pi t)^{-\frac d2} e^{-t\|\nabla g\|_{L^2(\R^d)}^2} \int_{\R^d} e^{-t (g^2*V)(x)}\,dx \,.
\end{align*}
Similarly as in the proof of the upper bound, we introduce spherical coordinates $x=r\omega$ and change variables $s=tr^\gamma$ to obtain
\begin{align*}
	\int_{\R^d}  e^{-t (g^2*V)(x)}\,dx & = \int_{\Sph^{d-1}} \int_0^\infty  e^{-t(g^2*V)(r\omega)} r^{d-1} \,dr\,d\omega\\
	& = \gamma^{-1} t^{-\frac d\gamma} \int_{\Sph^{d-1}} \int_0^\infty  e^{- t(g^2*V)((s/t)^\frac1\gamma \omega)} s^{\frac d\gamma-1} \,ds\,d\omega\,.
\end{align*}
It follows from Fatou's lemma that
\begin{equation}\label{eq: Fatou}
	\liminf_{t\to 0} t^{\frac d2 + \frac d\gamma} \Tr e^{-tH} \geq (4\pi)^{-\frac d2} \gamma^{-1}  \int_{\Sph^{d-1}} \int_0^\infty  \liminf_{t\to 0} e^{- t(g^2*V)((s/t)^\frac1\gamma \omega)}  s^{\frac d\gamma-1} \,ds\,d\omega \,.
\end{equation}
It follows by Lebesgue's differentiation theorem that, for a.e.\ $\omega\in\Sph^{d-1}$,
\begin{align*}
	\lim_{\epsilon \to 0}\epsilon^\gamma (g^2* V)( \epsilon^{-1} \omega) 
	& = \lim_{\epsilon \to 0}\int_{\R^d} \epsilon^{-d} g^2((\omega-x')/\epsilon) \ \epsilon^\gamma V(x'/\epsilon)\,dx'\\
	&= \lim_{\epsilon \to 0}\int_{\R^d}  \epsilon^{-d} g^2((\omega-x')/\epsilon) \ V(x')\,dx' \\
	&= V(\omega) \,.
\end{align*}
(Note that due to the fact that $g$ is symmetric decreasing, by the layer cake formula \cite{LiLo} the convolution with $g^2$ can be written as a superposition of convolutions with characteristic functions of balls and therefore Lebesgue's differentiation theorem is applicable. Initially, this theorem gives convergence for a.e.~$x\in\R^d$, but since $V$ is homogeneous this implies convergence for a.e.~$\omega$ with respect to surface measure on $\Sph^{d-1}$.) Setting $\epsilon =(t/s)^\frac 1\gamma$, we deduce that
$$
	\int_0^\infty  \liminf_{t\to 0} e^{- t(g^2*V)((s/t)^\frac1\gamma \omega)}  s^{\frac d\gamma-1} \,ds= \int_0^\infty  e^{- s V(\omega)} s^{\frac d\gamma-1} \,ds = \Gamma(\tfrac d\gamma)\, V(\omega)^{-\frac d\gamma} \,.
$$
When inserted in \eqref{eq: Fatou}, this completes the proof of the claimed lower bound. 

\begin{remark}\label{infiniterem}
	We claimed that for potentials of the form considered in Theorem \ref{main} the integral in Theorem \ref{warmup} is infinite. Let us justify this. Consider $F\in L^1(\Sph^{m-1}\times\Sph^{n-1})$ such that $V(x,y) = |x|^\alpha |y|^\beta F(x/|x|,y/|y|)$. We parametrize $\omega\in\Sph^{m+n-1}\subset\R^m\times\R^n$ as $(\Omega\sin\phi,\Theta\cos\phi)$ with $\Omega\in\Sph^{m-1}$, $\Theta\in\Sph^{n-1}$ and $\phi\in[0,\frac\pi 2]$. For the corresponding surface measure, we have $d\omega =  (\sin\phi)^{m-1} (\cos\phi)^{n-1}\,d\Omega\,d\Theta\,d\phi$ 
	and consequently
	$$
		\int_{\Sph^{m+n-1}}  V(\omega)^{-\frac{m+n}{\alpha+\beta}} \,d\omega
		= c \iint_{\Sph^{m-1}\times\Sph^{n-1}}  F(\Omega,\Theta)^{-\frac{m+n}{\alpha+\beta}}\,  d\Omega\,d\Theta
	$$
	with
	$$
		c = \int_0^{\frac\pi 2} (\sin\phi)^{m-1-\frac{(m+n)\alpha}{\alpha+\beta}}\, (\cos\phi)^{n-1-\frac{(m+n)\beta}{\alpha+\beta}} \,d\phi\,.
	$$
	The claim now follows from the fact that $c=\infty$ for the range of parameters in Theorem~\ref{main}. Indeed, if $\frac{m}{\alpha}\geq\frac n\beta$ (resp.~$\frac{m}{\alpha}\leq\frac n\beta$), then the integral diverges at $\phi=\frac\pi2$ (resp.~$\phi=0$). 
\end{remark}

%%%%%%%%%%%%%%%%%%%%%%%

\section{Partially semiclassical limit: Proof of Theorem \ref{mainheat}}

We now turn to the main application of our new inequality \eqref{eq:jensenbipartintro}, namely the proof of Theorem \ref{mainheat} (and thereby also that of Theorem \ref{main}).

As in the previous section we proceed by proving an upper and a lower bound on $\Tr e^{-tH}$. The upper bound is already contained in Simon's paper \cite{Si}, but we repeat the short argument to emphasize the similarity of the upper and lower bounds.

\medskip

For the \emph{upper bound} we apply the Golden--Thompson inequality, separating $-\Delta_x$ from the rest of the operator $H$. This is what Simon calls the `sliced Golden--Thompson inequality'. We obtain
\begin{align*}
	\Tr e^{-tH} & \leq \iint_{\R^m\times\R^m}   \Tr_{L^2(\R^n)}\bigl(e^{-t(|\xi|^2 -\Delta_y + V(x,y))}\bigr) \,\frac{dx\,d\xi}{(2\pi)^m}  \\
	& = (4\pi t)^{-\frac m2} \int_{\R^m} \Tr_{L^2(\R^n)}\bigl(e^{-t(-\Delta_y + V(x,y))}\bigr)\,dx \,.
\end{align*}
Introducing spherical coordinates $x=r\omega$ with $r>0$, $\omega\in\Sph^{m-1}$, we obtain
$$
	\int_{\R^m}  \Tr_{L^2(\R^n)}\bigl(e^{-t(-\Delta_y + V(x,y))}\bigr)\,dx = \int_{\Sph^{m-1}}\int_0^\infty  \Tr_{L^2(\R^n)}\bigl(e^{-t(-\Delta_y + r^\alpha V(\omega,y))}\bigr) r^{m-1}\,dr \, d\omega\,.
$$
Changing variables $y = g^{-\frac1{\beta+2}} y'$, we see that $-\Delta_y + r^\alpha V(\omega,y)$ is unitarily equivalent to $g^{\frac2{\beta+2}}(-\Delta_{y'} + r^\alpha g^{-1} V(\omega,y'))$ in $L^2(\R^n)$. We apply this observation with $g=r^\alpha$ and find
$$
\Tr_{L^2(\R^n)} \bigl(e^{-t(-\Delta_y + r^\alpha V(\omega,y))}\bigr) = \Tr_{L^2(\R^n)} \Bigl(e^{-tr^{\frac{2\alpha}{\beta+2}}K_\omega}\Bigr)
$$
with the operator $K_\omega := -\Delta_{y'} + V(\omega,y')$ in $L^2(\R^n)$. Changing variables $s=t r^\frac{2\alpha}{\beta+2}$ we obtain
\begin{align*}
	\int_0^\infty \!\!\Tr_{L^2(\R^n)} \bigl(e^{-t(-\Delta_y + r^\alpha V(\omega,y))}\bigr)r^{m-1}\,dr
	& = \int_0^\infty  \!\!\Tr_{L^2(\R^n)} \Bigl(e^{-tr^{\frac{2\alpha}{\beta+2}}K_\omega}\Bigr) r^{m-1}\,dr \\
	& = \frac{\beta+2}{2\alpha} t^{-\frac{m(\beta+2)}{2\alpha}} \!\int_0^\infty \!\! \Tr_{L^2(\R^n)}\bigl(e^{-s K_\omega}\bigr)s^{\frac{m(\beta+2)}{2\alpha}-1}\, ds \\
	& = \frac{\beta+2}{2\alpha} t^{-\frac{m(\beta+2)}{2\alpha}} \Gamma(\tfrac{m(\beta+2)}{2\alpha}) \, \Tr_{L^2(\R^n)} \Bigl(K_\omega^{-\frac{m(\beta+2)}{2\alpha}}\Bigr)\, .
\end{align*}
To summarize, we have shown that
$$
	\Tr e^{-tH} \leq C_{\frac{2\alpha}{\beta+2},m}' \, t^{- \frac{m(\alpha+\beta+2)}{2\alpha}} \int_{\Sph^{m-1}}  \Tr_{L^2(\R^n)}\Bigl(K_\omega^{-\frac{m(\beta+2)}{2\alpha}}\Bigr) \,d\omega\,,
$$
which is the claimed upper bound.

\medskip

We now turn to the proof of the \emph{lower bound}. We fix a symmetric decreasing function $g\in H^1(\R^m)$ with $\|g\|_{L^2(\R^m)} = 1$ and set, for $(x,\xi)\in\R^m\times\R^m$,
$$
\psi_{\xi,x}(x') = e^{i\xi\cdot x}\, g(x'-x) \,.
$$
Then, as is well known and can be deduced by Plancherel's theorem,
$$
\int_{\R^m\times\R^m}  |\psi_{\xi,x}\rangle \langle \psi_{\xi,x}| \,\frac{dx\,d\xi}{(2\pi)^m} = \1_{L^2(\R^m)} \,.
$$
Thus,
$$
\Tr e^{-tH} = 
\iint_{\R^m\times\R^m}  \Tr_{L^2(\R^m)} \left( |\psi_{\xi,x}\rangle \langle \psi_{\xi,x}| \Tr_{L^2(\R^n)} e^{-tH} \right)\,\frac{dx\,d\xi}{(2\pi)^m} \,.
$$
We now apply Theorem \ref{mainjensen} with $\phi(E) = e^{-tE}$ and $\rho = |\psi_{\xi,x}\rangle \langle \psi_{\xi,x}|$ (see also Remark \ref{mainjensenrem}) and obtain
$$
\Tr_{L^2(\R^m)} \bigl( |\psi_{\xi,x}\rangle \langle \psi_{\xi,x}| \Tr_{L^2(\R^n)} e^{-tH} \bigr)
\geq \Tr_{L^2(\R^n)} e^{-t \langle \psi_{\xi,x} | H | \psi_{\xi,x} \rangle} \,.
$$
Note that $\langle \psi_{\xi,x} | H | \psi_{\xi,x} \rangle$ is an operator in $L^2(\R^n)$. Using standard computations with coherent states we find
$$
\langle \psi_{\xi,x} | H | \psi_{\xi,x} \rangle = -\Delta_y + |\xi|^2 + \|\nabla g\|_{L^2(\R^m)}^2 + \widetilde V(x,y) \,,
$$
where
$$
\widetilde V(x,y) := \int_{\R^m} g(x-x')^2 V(x',y)\,dx' \,.
$$
Thus, we have shown that
\begin{align*}
	\Tr e^{-tH} & \geq \iint_{\R^m\times\R^m}  \Tr_{L^2(\R^n)}\bigl(e^{-t(-\Delta_y + |\xi|^2 + \|\nabla g\|_{L^2(\R^m)}^2 + \widetilde V(x,y))}\bigr) \,\frac{dx\,d\xi}{(2\pi)^m} \\
	& = (4\pi t)^{-\frac m2} e^{-t \|\nabla g\|_{L^2(\R^m)}^2} \int_{\R^m} \Tr_{L^2(\R^n)}\bigl(e^{-t(-\Delta_y + \widetilde V(x,y))}\bigr)\,dx \,.
\end{align*}
We now proceed similarly as in the upper bound. We introduce spherical coordinates $x=r\omega$, change variables by letting $y=r^{-\frac{\alpha}{\beta+2}} y'$ and $s=t r^\frac{2\alpha}{\beta+2}$. This gives
\begin{align*}
	\int_{\R^m}  &\Tr_{L^2(\R^n)} e^{-t(-\Delta_y + \widetilde V(x,y))}\,dx\\
	& = \int_{\Sph^{m-1}} \int_0^\infty  \Tr_{L^2(\R^n)}\bigl(e^{-t(-\Delta_y + \widetilde V(r\omega,y))}\bigr) r^{m-1}\,dr \, d\omega\\
	& = \int_{\Sph^{m-1}} \int_0^\infty  \Tr_{L^2(\R^n)}\Bigl(e^{-tr^\frac{2\alpha}{\beta+2}(-\Delta_{y'} + r^{-\alpha} \widetilde V(r\omega,y'))}\Bigr)r^{m-1}\,dr \, d\omega \\
	& = \frac{\beta+2}{2\alpha} \, t^{-\frac{m(\beta+2)}{2\alpha}} \int_{\Sph^{m-1}} \int_0^\infty \Tr_{L^2(\R^n)}\Bigl(e^{-s K_{\omega}^{(\epsilon_{s,t})}}\Bigr)s^{\frac{m(\beta+2)}{2\alpha}-1}\,ds\,d\omega \,,
\end{align*}
where $\epsilon_{s,t}:= (t/s)^{\frac{\beta+2}{2\alpha}}$ and
$$
K_{\omega}^{(\epsilon)} := -\Delta_{y'} + \epsilon^\alpha \widetilde V(\epsilon^{-1} \omega,y')
\qquad\text{in}\ L^2(\R^n) \,.
$$
We shall show below that for a.e.~$\omega\in\Sph^{m-1}$ and every $s>0$
\begin{equation}
	\label{eq:mainheatproof}
	\liminf_{\epsilon\to 0} \Tr_{L^2(\R^n)} e^{-s K_{\omega}^{(\epsilon)}} \geq \Tr e^{-s K_\omega} 
\end{equation}
with the same operator $K_\omega$ as in the upper bound. Therefore, by Fatou's lemma,
\begin{align*}
	\liminf_{t\to 0} t^{\frac{m(\alpha+\beta+2)}{2\alpha}} \Tr e^{-tH} 
	& \geq \frac{\beta+2}{2\alpha(4\pi)^{\frac m2}}  \int_{\Sph^{m-1}} \!\int_0^\infty  \!\liminf_{t\to 0} \Tr_{L^2(\R^n)}\Bigl(e^{-s K_{\omega}^{(\epsilon_{s,t})}}\Bigr)  s^{\frac{m(\beta+2)}{2\alpha}-1}\,ds\,d\omega\\
	& \geq  \frac{\beta+2}{2\alpha(4\pi)^{\frac m2}}  \int_{\Sph^{m-1}} \!\int_0^\infty \! \Tr_{L^2(\R^n)}\bigl(e^{-s K_\omega}\bigr)  s^{\frac{m(\beta+2)}{2\alpha}-1}\,ds\,d\omega\\
	& = C_{\frac{2\alpha}{\beta+2},m}' \int_{\Sph^{m-1}}  \Tr_{L^2(\R^n)} \Bigl(K_\omega^{-\frac{m(\beta+2)}{2\alpha}}\Bigr) \,d\omega \,.
\end{align*}
This is the claimed lower bound.

It remains to prove \eqref{eq:mainheatproof}. To this end we use the following lemma.

\begin{lemma}\label{lsc}
	Let $A_j$, $j\in\N$, and $A$ be selfadjoint, lower bounded operators in a Hilbert space with a common form core $\mathcal Q$ and assume that for all $\psi\in\mathcal Q$ we have $\limsup_{j\to\infty} \langle \psi | A_j |\psi \rangle \leq \langle \psi | A |\psi \rangle$. Then
	$$
	\liminf_{j\to\infty} \Tr e^{-A_j} \geq \Tr e^{-A} \,.
	$$
\end{lemma}

The following proof of the lemma relies on the Gibbs variational principle (see, e.g., \cite[Theorem 7.45]{Car2}), which says that for any selfadjoint, lower semibounded operator~$H$
$$
\inf_{\rho \ \text{density matrix}} \left( \Tr \rho^\frac12 H \rho^\frac12 + \Tr \rho\ln\rho \right) = - \ln \Tr e^{-H} \,,
$$
where the infimum is taken over all density matrices with $\Tr \rho\ln\rho>-\infty$. Note that $\Tr \rho^\frac12 H \rho^\frac12$ is well defined, but possibly $+\infty$.

\begin{proof}
	Let $\rho$ be a finite rank density matrix with range in $\mathcal Q$. Then, by assumption,
	$$
	\Tr \rho^\frac12 A \rho^\frac12 + \Tr \rho\ln\rho \geq \limsup_{j\to\infty} \left( \Tr \rho^\frac12 A_j \rho^\frac12 + \Tr \rho\ln\rho \right).
	$$
	Bounding the right side from below by the Gibbs variational principle, we find
	$$
	\Tr \rho^\frac12 A \rho^\frac12 + \Tr \rho\ln\rho \geq - \liminf_{j\to\infty} \ln \Tr e^{-A_j} \,.
	$$
	By density this lower bound extends to any density matrix $\rho$ with $\Tr \rho\ln\rho>-\infty$. Taking the infimum over all such $\rho$ and employing again the Gibbs variational principle, we arrive at
	$$
	- \ln \Tr e^{-A} \geq - \liminf_{j\to\infty} \ln \Tr e^{-A_j} \,,
	$$
	which is the claimed inequality.
\end{proof}

We return to the proof of \eqref{eq:mainheatproof}. By Fubini's theorem and homogeneity of $V$ there is a full measure subset of $\Sph^{m-1}$ such that for any $\omega$ from this set and any $\epsilon>0$ the function $y'\mapsto \epsilon^\alpha \widetilde V(\epsilon^{-1}\omega,y')$ is locally integrable on $\R^n$. Since it is also nonnegative, it follows that $C^\infty_c(\R^n)$ is a form core for $K_\omega^{(\epsilon)}$; see, e.g., \cite[Proposition 4.1]{FrLaWe}. We claim that for $\psi\in C^\infty_c(\R^n)$ we have
\begin{equation}
	\label{eq:mainheatproof2}
	\lim_{\epsilon \to 0}\langle \psi | \epsilon^\alpha \widetilde V(\epsilon^{-1}\omega,\cdot) |\psi \rangle = \langle \psi | V(\omega,\cdot) |\psi \rangle \,.
\end{equation}
Once we have shown \eqref{eq:mainheatproof2}, we can apply Lemma \ref{lsc} with $A_j = s K_\omega^{(\epsilon_j)}$ and obtain \eqref{eq:mainheatproof}.

To prove \eqref{eq:mainheatproof2}, we set $W(x) := \int_{\R^n} V(x,y) |\psi(y)|^2\,dy$ and note that \eqref{eq:mainheatproof2} is equivalent to
$$
\lim_{\epsilon \to 0}\epsilon^\alpha (g^2*W)(\epsilon^{-1}\omega) = W(\omega) \,.
$$
This holds for a.e.~$\omega\in\Sph^{m-1}$ as shown in the proof of the lower bound in Theorem \ref{warmupheat}. This concludes the proof.

%%%%%%%%%%%%%%%%%%%%%%%%%%%%%%%%%%%%%%%%%%%%%%%%%%%%%%%%%%%%%%%%%%%%%%%%%%%%%%%%
%%%%%%%%%%%

\bibliographystyle{amsalpha}

\begin{thebibliography}{22}
	
\bibitem{AlLa} A. Aljahili, A. Laptev, \textit{Non-classical spectral bounds for Schr\"odinger operators}. J. Math. Sci. (N.Y.) \textbf{270} (2023), no. 6, Problems in mathematical analysis. No. 124, 741--751.	
	
\bibitem{Be} F. A. Berezin, \textit{Covariant and contravariant symbols of operators}. Izv. Akad. Nauk SSSR Ser. Mat. \textbf{36} (1972), 1134--1167 (Russian); English translation in Math. USSR-Izv. \textbf{6} (1972), 1117--1151.

\bibitem{BiSo} M. Sh. Birman, M. Z. Solomjak, \textit{Quantitative analysis in Sobolev imbedding theorems and applications to spectral theory}. American Mathematical Society Translations, Series 2, 114. American Mathematical Society, Providence, RI, 1980.

\bibitem{Ca} T. Carleman, \textit{Propri\'et\'es asymptotiques des fonctions fondamentales des membranes vibrantes}. Attonde skand. matematikerkongressen i Stockholm 1934, 34--44.

\bibitem{Car1} E. A. Carlen, \textit{Trace inequalities and quantum entropy: an introductory course}. Contemp. Math., 529. American Mathematical Society, Providence, RI, 2010, 73--140.

\bibitem{Car2} E. A. Carlen, \textit{Inequalities in matrix algebras}. Book in preparation.

\bibitem{CdVdHDiTr} Y. Colin de Verdi\'ere, C. Dietze, M. V. de Hoop, E. Tr\'elat, \textit{Weyl formulae for some singular metrics with application to acoustic modes in gas giants}. Preprint (2024), arXiv:2406.19734.

\bibitem{DiRe} C. Dietze, L. Read, \textit{Concentration of eigenfunctions on singular Riemannian manifolds}. Preprint (2024), arXiv:2410.20563.

\bibitem{Fr} R. L. Frank, \textit{Weyl's law under minimal assumptions}. In: From complex analysis to operator theory--a panorama, 549--572, Oper. Theory Adv. Appl., 291, Birkh\"auser/Springer, Cham, 2023.

\bibitem{FrLaWe} R. L. Frank, A. Laptev, T. Weidl, \textit{Schr\"odinger operators: eigenvalues and Lieb--Thirring inequalities}, Cambridge Studies in Advanced
Mathematics, Cambridge University Press, Cambridge, 2023.

\bibitem{FrLa} R. L. Frank, S. Larson, \textit{Semiclassical inequalities for Dirichlet and Neumann Laplacians on convex domains}. Preprint (2024), arXiv:2410.04769.

\bibitem{LaWe} A. Laptev, T. Weidl, \textit{Sharp Lieb--Thirring inequalities in high dimensions}. Acta Math. \textbf{184} (2000), no. 1, 87--111.

\bibitem{Li} E. H. Lieb, \textit{The classical limit of quantum spin systems}. Comm. Math. Phys. \textbf{31} (1973), 327--340.

\bibitem{Li2} E. H. Lieb, \textit{Coherent states as a tool for obtaining rigorous bounds}. In: Coherent States: Past, Present, and Future, eds. D.H. Feng, J.R. Klauder, and M.R. Strayer, World Scientific, Singapore, 1994.

\bibitem{LiLo} E. H. Lieb, M. Loss, \textit{Analysis}. Second edition. Grad.~Stud.~Math., 14, American Mathematical Society, Providence, RI, 2001.

\bibitem{Re} L. Read, \textit{On the asymptotic number of low-lying states in the two-dimensional confined Stark effect}. Preprint (2024), arXiv:2404.14363.

\bibitem{Ro} D. Robert, \textit{Comportement asymptotique des valeurs propres d'op\'erateurs du type Schr\"odinger \`a potentiel ``d\'eg\'en\'er\'e''}. (French) J. Math. Pures Appl. (9) \textbf{61} (1982), no. 3, 275--300.

\bibitem{Si0} B. Simon, \textit{Functional integration and quantum physics}. Second edition. AMS Chelsea Publishing, Providence, RI, 2005.

\bibitem{Si1} B. Simon, \textit{The classical limit of quantum partition functions}. Comm. Math. Phys. \textbf{71} (1980), no. 3, 247--276.

\bibitem{Si} B. Simon, \textit{Nonclassical eigenvalue asymptotics}. J. Funct. Anal. \textbf{53} (1983), no. 1, 84--98. 

\bibitem{So} M. Solomyak, \textit{Asymptotic behavior of the spectrum of the Schr\"odinger operator with nonregular homogeneous potential}. Mat. Sb. (N.S.) \textbf{127}(169) (1985), no. 1, 21--39, 142 (Russian); English translation in Math. USSR-Sb. \textbf{55} (1986), no. 1, 19--37.

\bibitem{Ta} H. Tamura, \textit{The asymptotic distribution of eigenvalues of the Laplace operator in an unbounded domain}. Nagoya Math. J. \textbf{60} (1976), 7--33.
	
\end{thebibliography}

\end{document}